\documentclass[preprint,showpacs,preprintnumbers,amsmath,amssymb]{revtex4-1}

\usepackage{graphicx}
\usepackage{dcolumn}
\usepackage{bm}
\usepackage{booktabs}
\usepackage{amsmath}
\usepackage{multirow}

\newcommand{\br}{\mathbf{r}}

\begin{document}
\title{A van der Waals density functional for solids}
\author{Torbj\"orn Bj\"orkman}
 \affiliation{COMP Centre of Excellence, Department of Applied Physics, Aalto University School of Science, P.O. Box 11100, 00076 Aalto, Finland}
\date{\today}

\begin{abstract}
The recent non-local correlation functional of Vydrov and van Voorhis[J. Chem. Phys. {\bf133}, 244103 (2010)] is investigated and two new versions of the functional are suggested as being appropriate for describing van der Waals interactions in solids. A refitting of the original functional is demonstrated to result in very accurate interlayer binding energies for weakly bonded layered solids.  A VV10 functional based on the generalized gradient approximation by Armiento and Mattsson[Phys. Rev. B {\bf72}, 085108 (2005)], while performing slightly worse for interlayer binding is highly successful in describing the equilibrium geometries of both weakly bonded and close packed solids. 
\end{abstract}
\maketitle

The development of non-local exchange-correlation (XC) functionals that allow for a description of van der Waals (vdW) interaction, an effect in principle not describable by local or semilocal approximations, is a significant recent advance in density functional theory. The field was pioneered by a Rutgers/Chalmers collaboration that developed the basic framework over a number of years, first leading up to a functional for layered solids\cite{andersson1996,hult1996,rydberg2000,rydberg2003} which was generalized to general geometries\cite{llvdw1} and which has since then been further improved\cite{llvdw2}.
Within this framework, Vydrov and van Voorhis (VV) have in a series of papers elaborated on the original underlying local polarizability model\cite{vydrov2009a,vydrov2009b,vydrov2010a} and suggested the functional VV10\cite{vv10} derived by fitting of interaction energies to a set of molecules. While part of the earlier work of VV has been criticized\cite{LLhateVV,VVreply}, VV10 was nevertheless shown to be highly accurate for molecules and for geometries in weakly bonded solids\cite{bjorkman2012a,bjorkman2012b}. However, it was found that VV10 greatly overestimates the binding energy in weakly bonded solids compared to more precise methods\cite{bjorkman2012a,bjorkman2012b}, indicating that the original parameters selected fitting for molecular interaction energies are unsuitable for solids. In fact, of the various flavors of vdW corrected methods investigated in Refs. \onlinecite{bjorkman2012a,bjorkman2012b}, none was entirely successful in producing uniformly reliable results for weakly bonded layered solids, providing a strong motivation for improving on the vdW methodology also in solids. The need to bias functionals towards either molecules or solids is common, most famously the generalized-gradient approximation (GGA) form will tend to carry a bias either way, due to lack of flexibility in the functional form\cite{pbesoldebate1,pbesoldebate2}. In the case of vdW interactions, recent results show that a method that is successful for molecules can, by inclusion of self-screening effects be greatly improved, both when applied for molecules\cite{tkatchenko2012} and for adsorption energies\cite{ruiz2012}. In solids, the screening is a large effect and so it is not surprising that VV10, a method fitted for molecules, tends to overestimate the interactions.
In this Brief Report, I will, rather than explicitly modeling the screening, simply refit VV10 for solids, using a suitable parent functional. I perform an investigation of the performance of the GGA functionals PW86R\cite{murray2009}, AM05\cite{am05} and PBEsol\cite{pbesol}, followed by the fitting of VV10 for weakly bonded layered solids using PW86R and AM05 as parent functionals. Apart from the layered solids used in the fitting procedure, the functionals are tested for 23 non-vdW-bonded solids\cite{pbesol} and for the S22 training set of molecules\cite{s22}.

In the formalism of Refs. \onlinecite{llvdw1,vv10} the correlation energy is split as
\begin{equation}
E_{c} = E^{0}_{c} + E_c^{nl},
\end{equation}
where $E^{0}$ denotes the local part of the correlation and $E^{nl}_c$ is a non-local part in the form
\begin{equation}
E^{nl}_c = \frac{\hbar}{2} \int\int d\br d\br' n(\br) \Phi(\br,\br') n(\br'),
\end{equation}
where $n(\br)$ is the electron density at $\br$ and $\Phi(\br,\br')$ is a function that describes the density-density interaction. 
In the original formalism\cite{llvdw1,llvdw2}, the exchange energy was taken from a GGA and $E^{0}_c$ from the local density approximation (LDA) and the GGA has been chosen to represent the exchange energy well in sparse systems either by selection of an appropriate GGA\cite{llvdw1,llvdw2,vv10} or by refitting the exchange to some set of systems\cite{klimes2010}. VV10 instead uses the full parent functional also for the local correlation, thus effectively turning the non-local correlation part into an additional correction to be applied on top of the parent functional. The local response model of VV10 and the function $\Phi$ are described in detail in Ref. \onlinecite{vv10}, and here we just note that it depends on the density through a local response parameter, $\omega_0(\br)$ related to the local plasma frequency, $\omega_p(\br)$, and that it also contains a parameter $\kappa(\br) = b\frac{v_F^2(\br)}{\omega_p(\br)}$, where $v_F(\br)$ is the local fermi velocity, that controls the short range damping of the vdW contribution and a local band-gap, $\omega_g(\br) \propto C\left|\frac{\nabla n}{n}\right|^4$. The local band-gap was introduced to keep the static polarizability from diverging without having to introduce an explicit integration cutoff and the value of $C$ determines the long range asymptotic behavior of the functional. The dependence of the energy on the parameters is not always straightforward due to the dependence on the density gradient, but generally speaking, larger values of either parameter tends to decrease $E^{nl}_c$.
The parameters $b$ and $C$ are to be determined for each parent functional by fitting them to some desirable property. VV determined $C$ by optimizing the $C_3$ coefficients for a set of atoms and molecules and subsequently determined $b$ to minimize the errors in the binding energies of the S22 training set\cite{s22}, and used as parent functionals the GGA PW86R\cite{murray2009}, developed to reproduce exchange properties so as to be suitable for use in vdW functionals\cite{llvdw2}, and to the range separated hybrid functional LC-$\omega$PBE\cite{vydrov2006}, developed to cure errors in the long range interaction induced by electron self-interaction.

As previously mentioned, recent studies have shown\cite{bjorkman2012a,bjorkman2012b} that, for weakly bonded layered compounds, the VV10 functional produces equilibrium geometries in fair agreement with experiment, with vdW bond lengths being only slightly overestimated, but that interlayer binding energies in comparison with RPA are consistently around 50\% too large. Since adjustment of parameters to decrease the large overshoot in the binding energy will make the already slightly too large vdW bond lengths deteriorate, a working solution for solids is requires changing the GGA functional which controls the repulsive part on the compression side of the binding energy curve.  Since in the VV10 framework, the vdW correction is applied on top of the full parent functional, it is not necessarily very important that the exchange part of the functional \emph{by itself} is accurately represented, but rather that the sum of exchange and correlation is accurate and has the expected behavior of a semilocal functional, i.e.  yields zero or very small binding for vdW dominated systems,  thus avoiding double counting of the interactions\cite{rydberg2003,vv10}. Based on previous results\cite{haas2009,ann_detonation}, the GGA functional AM05\cite{am05} is expected to have the desired property of little or no binding in vdW dominated systems. Its construction uses a fitting of the \emph{total} functional for a jellium surface, yielding a good combined description of the XC, despite being less accurate for exchange and correlation separately\cite{mattsson2010} and has been shown to performs very well for regular solids\cite{haas2009,mattsson2008}. The combination of the desirable properties of a good total XC functional with very small binding in vdW dominated systems makes it a good candidate for a parent functional to a vdW density functional for solids. Another possible candidate is the PBEsol functional\cite{pbesol}, which is somewhat related to AM05 in that they are both based on fits to a jellium surface, but where AM05 has been fitted for the full functional, in PBEsol first the exchange is fitted and then a compatible correlation is added. These  differences aside, AM05 and PBEsol show very similar performance for solids where vdW interactions are not important\cite{mattsson2008,haas2009} and thus both will be tested here.

All calculations were performed using the \textsc{vasp} code\cite{vasp3} with real space implementation of the non-local vdW functionals\cite{andrisvdw}. The same technical settings as those used in Ref. \onlinecite{bjorkman2012a} were used for the layered solids. For molecules in the S22 training set, a planewave cutoff of 400 eV and a cubic cell with sides of length 15\AA{} were used, and these settings that were verified to yield results very close to those obtained by VV\cite{vv10}. For the non-vdW-bonded solids, the planewave cutoff and $k$-space sampling were increased until the change in total energy was less than 1 meV and Brillouin zone integrations were performed using adaptive gaussian smearing\cite{ags}. As reference for the binding energies of the weakly bonded layered compounds, the direct random-phase approximation (RPA) data of Ref. \onlinecite{bjorkman2012a} for 26 layered solids was used, and geometrical properties were compared with experimental data, without accounting for zero-point anharmonic expansion (ZPAE) corrections. The layered solids with their experimental references were: BN\cite{Brager1937}, HfS$_2$\cite{McTaggart1958}, HfSe$_2$\cite{McTaggart1958}, HfTe$_2$\cite{Brattas1971}, MoS$_2$\cite{Py1983}, MoSe$_2$\cite{Bronsema1986}, MoTe$_2$\cite{Wyckoff1963}, NbSe$_2$\cite{Meerschaut2002}, NbTe$_2$\cite{Allakhverdiev1969}, PbO\cite{Boher1992}, PdTe$_2$\cite{pell1996}, PtS$_2$\cite{Soled1976}, PtSe$_2$\cite{Furuseth1965}, TaS$_2$\cite{Spijkerman1992}, TaSe$_2$\cite{Bjerkelund1967}, TiS$_2$\cite{Kusawake2000}, TiSe$_2$\cite{Minagawa1984}, TiTe$_2$\cite{Arnaud1981}, VS$_2$\cite{Wiegers1980}, VSe$_2$\cite{Wiegers1980}, WS$_2$\cite{Schutte1987}, WSe$_2$\cite{Schutte1987}, ZrS$_2$\cite{McTaggart1958}, ZrSe$_2$\cite{Ahouandjinou1976}, ZrTe$_2$\cite{McTaggart1958} and graphite\cite{Wyckoff1963}. Calculations of binding energies were done with the intralayer geometry frozen and only the layer distance being varied, to conform to the settings used in the RPA calculations of Ref.~\cite{bjorkman2012a}. The calculations of equilibrium geometries were done by minimizing the total energy for a series of fixed volumes (to minimize errors from Pulay stress) while allowing for complete relaxation of internal positions and cell shape. The reference data for the S22 training set were taken from Jurecka et al.\cite{s22}, and for non-vdW-bonded solids, the 23 solids tested by Klime\v{s} et al. in Ref. \onlinecite{klimes2010}, including ZPAE corrections, were used. 

\begin{figure}[htbp] 
   \centering
   \includegraphics[width=\textwidth]{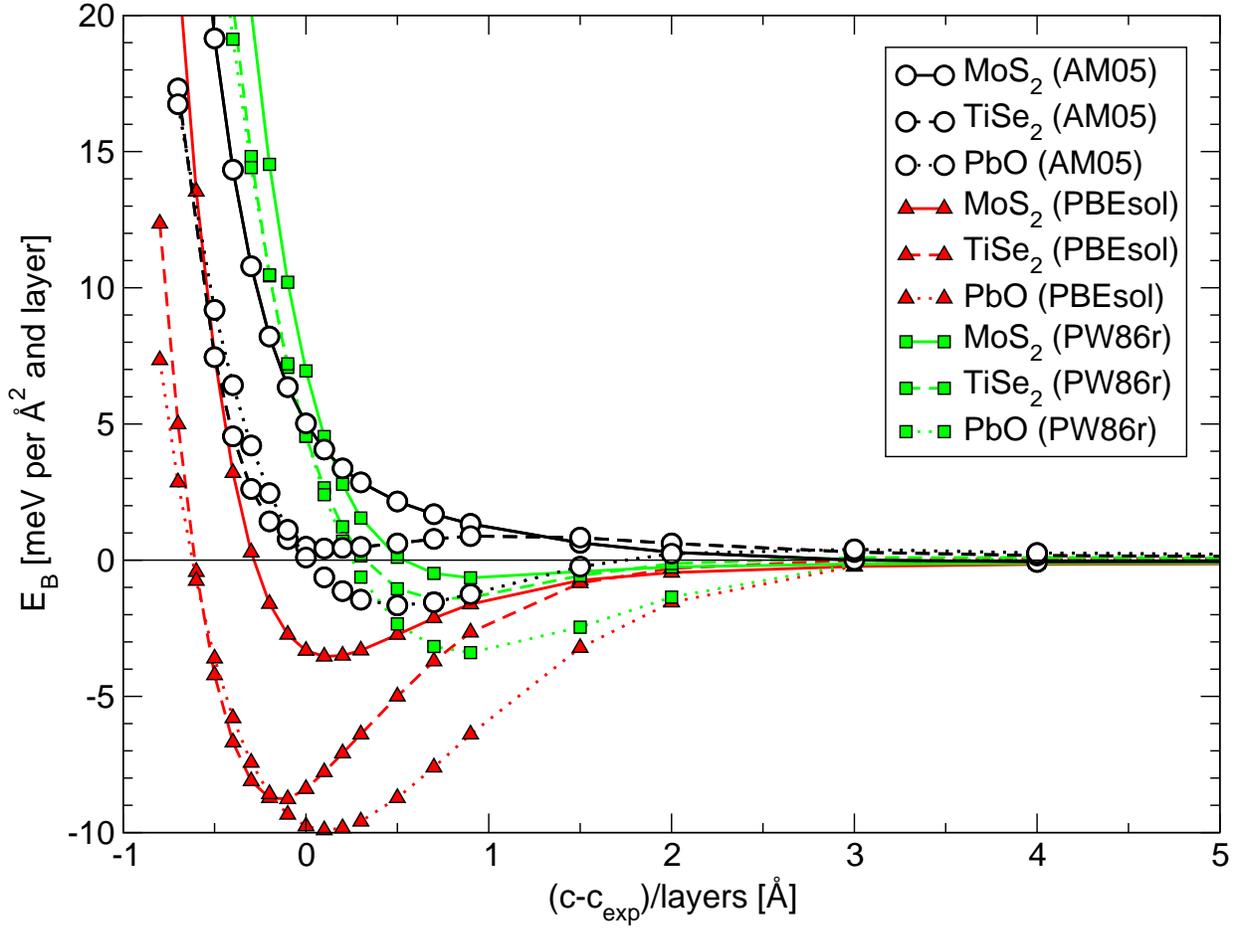}
   \caption{Binding energy curves for three representative compounds from the set used for optimizing the VV10 functional for solids, for the GGA functionals AM05, PBEsol and PW86R. The units of the $x$-axis are the deviation of the $c$ axis length from the experimental geometry, and the units of the $y$-axis are meV/\AA$^2$, both normalized by the number of layers per unit cell. The calculated values of $(c-c_{exp})/\text{layers}$ extend out to 15.0 \AA. In the examples shown here, AM05 has no binding at all for MoS$_2$ and TiSe$_2$, although for TiSe$_2$ there is a local minimum near the experimental geometry, whereas PbO has a small binding energy of about 1.5 meV/\AA$^2$. By contrast, PW86R always gives a small binding energy with the minimum in the vicinity of the experimental lattice constant and PBEsol gives a much deeper minimum close to the experimental lattice constant. Note also the lower slope of AM05 in the compressive region to the left of the experimental equilibrium lattice constant.}
   \label{E-c}
\end{figure}

To see how appropriate PW86R, AM05 and PBEsol are as parent functionals for a vdW density functional for solids, the 26 layered compounds were first investigated using the bare GGA functionals. The result is illustrated in Figure~\ref{E-c}, where a set of representative curves for the total energy as function of the $c$ axis length are shown\cite{supplemental}. AM05 consistently show the least binding, with only 7 out of 26 layered compounds having a global minimum at finite $c$ axis length, but in many cases a local minimum near the experimental equilibrium geometry is seen. With the exception of graphite, PW86R always has a global minimum in the vicinity of the experimental equilibrium geometry and the binding energy is always larger than that of AM05. By contrast, PBEsol always binds more strongly and with a much larger spread of the binding energies from 1.3 meV/\AA$^2$ for graphite to 14.5 meV/\AA$^2$ for TiTe$_2$, 7\% and 76\% of the RPA reference values, respectively\footnote{PdTe$_2$ has a significant binding energy with all three GGA functionals and is discussed in Ref.~\cite{bjorkman2012a}.}. Because of its large binding, PBEsol must be considered less well suited as a parent functional for VV10, whereas the original PW86R is much better, and AM05 even more so, and so I have  investigated only PW86R and AM05 as parent functionals for the refitted VV10 functional. Important to note here is also that the AM05 functional is clearly softer on the compression side than the PW86R functional, as can be seen by inspecting the slope of the curves in Figure~\ref{E-c}. Combining naming conventions from Refs. \onlinecite{vv10} and \onlinecite{pbesol}, the resulting functionals have been labelled PW86R-VV10sol and AM05-VV10sol.

The original VV10 functional had first the long-range behavior fitted to a set of $C_3$ coefficients by adjustment of the parameter $C$, and then interaction energies of the S22 set fitted using $b$. In analogy with this, I fit the $b$ parameter to the RPA binding energies of layered solids. The $C$ parameter is less straightforward, since, while the long-range behavior of the vdW interaction is similar for all finite fragments, following a $R^{-6}$ power law, the power laws for the distance dependence of the interaction between the infinite sheets of a layered solid depend on the electronic structure of the layers\cite{dobson2006}. Since the form of the VV10 functional (as well as all other vdW density functionals) are constructed to produce an $R^{-6}$ behavior at long distance, they will asymptotically follow a  $R^{-4}$ power law at large separation for two-dimensional sheets, irrespective of the electronic structure. To fit the long range behavior for solids, we would thus be forced to constrain the investigation to compounds with a gap, where $R^{-4}$ is the correct power law\cite{dobson2006}. Unfortunately, the only available high-level calculation reference data for the long-range behavior of the sheets of layered solids is for graphite\cite{seb}. This is obviously insufficient for a reliable fitting procedure and furthermore, graphite is disqualified by not having a band gap. Here, the lack of appropriate fitting data was resolved by keeping the original VV10 value of the $C$ parameter determined for molecules for PW86R and for AM05 we fit also $C$ to optimize the binding energies for layered solids. In this way, two new functionals were obtained: PW86R-VV10sol ($b=9.15$, $C=0.0093$) and AM05-VV10sol ($b=10.25$, $C=10^{-6}$). 

The value of $C$ optimized for the AM05 parent functional requires a remark. The minimization of the errors in binding energies proceeded in steps by first optimizing $b$ with $C=0.0089$, the value used in the original VV10 functional. Then fixing the $b$ parameter and varying $C$, it was found that decreasing $C$ to zero had the effect of improving almost all of the binding energies, irrespective of whether they were too high or too low. The $b$ parameter was then refitted once more, moving only slightly from its previous value, and this point was found to still be the minimum for $C$, which was not allowed to take on negative values, since this would yield unphysical negative values of $\omega_g$. Setting $C$ to zero, thus eliminating the local gap parameter $\omega_g$ will give a formal problem with the functional, since the gap parameter was introduced to keep the static polarizability from diverging. However, setting the value of $C$ to some small number, here $10^{-6}$ was chosen, will cure this formal problem, although it was noted that whether $C$ was set to be identically zero or a small number appeared to make no difference in practice.

\begin{table}[htdp]
\caption{Mean relative errors (MRE) and mean absolute relative errors (MARE) in percent for binding energies and lattice constants for 26 weakly bonded layered solids and 23 strongly bonded solids for the investigated functionals\cite{supplemental}. The optimized quantities for the different functionals are shown in bold font. The comparison of equilibrium geometries for the 23 solids include ZPAE corrections but for the 26 layered solids the published experimental data is used.}
\begin{center}
\begin{tabular}{l|cc|cc|cc|cc|cc}
\multicolumn{1}{l}{} & \multicolumn{6}{|c}{26 layered solids} & \multicolumn{2}{|c}{23 solids} & \multicolumn{2}{|c}{S22}\\
 &\multicolumn{2}{c}{$E_B$} & \multicolumn{2}{c}{$c$} & \multicolumn{2}{c}{$a$} & \multicolumn{2}{|c}{$a$} &\multicolumn{2}{|c}{$E_{int}$}\\
Functional & MRE & MARE & MRE & MARE & MRE & MARE & MRE & MARE & MRE & MARE\\\hline\hline
PW86R-VV10 & 52.5 & 52.5 & 0.7 & 1.1 & 1.8 & 1.8 & 0.5 & 1.7 & 2.6 & {\bf 4.9} \\
PW86R-VV10sol & 0.04 & {\bf 6.9} & 3.4 & 3.4 & 2.2 & 2.2 & 1.1 & 1.8 & -25.5 & 26.5 \\
AM05-VV10sol & 5.2 & {\bf 11.1} & -0.21& 1.6 & -1.2 & 1.4 & 0.0 & 0.7 & -34.2 & 36.3
\end{tabular}
\end{center}
\label{layertable}
\end{table}%

Table~\ref{layertable} shows the mean relative errors (MRE) and mean absolute relative errors (MARE) for the tests carried out for the different functionals. The original VV10 functional (here labelled PW86R-VV10) and it is clear that its performance for the molecular interaction energies ($E_{int}$) of the S22 set is superior to the functionals fitted for solids, while giving much too high binding energies for the layered solids. PW86R-VV10sol achieves small errors for the binding energies of the layered solids but yields rather large lattice constants both for the layered and non-layered solids. In the comparison of equilibrium geometries the role of the ZPAE corrections needs to be considered. For the set of 23 regular solids, ZPAE corrected reference data is available\cite{klimes2010}, but for the lattice constants of weakly bonded solids no such data is available, since the standard way of estimating ZPAE corrections are based on reliable first principles calculations\cite{alchagirov2001}. For graphite, the ZPAE expansion of the $c$ axis length has previously been estimated as high as 0.5\%\cite{kelly1975}, which would put the AM05-VV10sol values in excellent agreement with experiment. Also both the in-plane lattice constant and the lattice constants for regular solids are clearly better for AM05-VV10sol than both PW86R-VV10 and PW86R-VV10sol, which reflects the softer behavior of AM05 on the compression side. To improve on the geometries for PW86R-VV10sol we would need to increase the vdW interaction component, which would lead to overestimation of the binding energies. 

The results of Table \ref{layertable} shows that the methodology of VV is clearly fails to simultaneously capture the binding characteristics of molecules and solids. Given the rather similar behavior for the two different parent functionals, PW86R and AM05, when refitted for solids, it appears unlikely that this is attributable to problems with the parent functional, but must be ascribed to the local polarizability model itself. The performance for molecules of the original functional is excellent, and for weakly bonded layered systems the results are almost equally good for the refitted versions. This is a substantial improvement over the results obtained by Bj\"orkman et al. in Refs. \onlinecite{bjorkman2012a,bjorkman2012b}, where either the interlayer distance or the binding energy was found to be too large for the vdW density functionals. In view of the high interest in single layer graphene and h-BN systems, the good performance of AM05-VV10sol for graphite and h-BN should be pointed out, with excellent agreement with experiment for in-plane lattice parameters (errors of 0.2\% for graphite and -0.1\% for h-BN) and good agreement for the interlayer distance (errors of 4.4\% for graphite and 2.2\% for h-BN) and interlayer binding energy (errors of -6.1\% for graphite and 4.6\% for h-BN). The better equilibrium geometries obviously makes AM05-VV10sol the most appropriate method for obtaining accurate geometries for solids, but this more appropriate balance of the vdW component to the parent GGA leads the present author to suggest that it is the more appropriate choice for most purposes regarding solid state calculations where vdW interactions plays a role. 

The author wishes to thank A.~Gulans for stimulating and enlightening discussions and R.~M.~Nieminen for constructive criticism on the manuscript. This research was supported by the Academy of Finland through the COMP Centre of Excellence Grant 2012-2017. Computational resources were provided by Finland's IT center for Science (CSC).

%
\end{document}